\documentclass[10pt]{article}
\usepackage{natbib}
\usepackage{graphicx}
\usepackage{eurosym}
\usepackage[english]{babel}

\usepackage{amssymb,amsmath}
\usepackage{natbib}
\usepackage{color}
\usepackage{placeins} % Provide the FloatBarrier command 
\usepackage{multirow}
\usepackage[left]{lineno}
\usepackage{setspace}
\usepackage{ulem}

\DeclareGraphicsExtensions{.png}
\usepackage[T1]{fontenc}

\begin{document}

\definecolor{gris25}{gray}{0.20}

%\begin{frontmatter}
\title{EvaSylv: A user-friendly software to evaluate forestry scenarii including natural risk}
\author{Patrice Loisel\thanks{MISTEA, INRA, Montpellier SupAgro, Univ Montpellier, Montpellier, France}, Guillerme Duvilli\'e\thanks{GOM, Univ Libre de Bruxelles, Belgique}, Denis Barbeau\thanks{MISTEA, INRA, Montpellier SupAgro, Univ Montpellier, Montpellier, France}, Brigitte Charnomordic\thanks{MISTEA, INRA, Montpellier SupAgro, Univ Montpellier, Montpellier, France}} 
\date{}

\maketitle{}
\begin{spacing}{1.5}

\begin{abstract}
\begin{itemize}
\item {\bf Context} Forest management relies on the evaluation of silviculture practices. The  increase in natural risk due to climate change makes it necessary to consider evaluation criteria that take natural risk into account. Risk integration in existing software requires advanced programming skills.
\item {\bf Aims} We propose a user-friendly software to simulate even-aged and monospecific forest at the stand level, in order to evaluate and optimize forest management. The software gives the possibility to run management scenarii with or without considering the impact of natural risk. The control variables are the dates and rates of thinning and the cutting age.
\item {\bf Methods} The risk model is based on a Poisson processus. The Faustmann approach, including tree damage risk, is used to evaluate future benefits, economic or ecosystem services. It relies on the calculation of expected values, for which a dedicated mathematical development has been done. The optimized criteria used to evaluate the various scenarii are the Faustmann value and the Averaged yield value.
\item {\bf Results} We illustrate the approach and the software  on two case studies: economic optimization of a beech stand and carbon sequestration optimization of a pine stand.
\item {\bf Conclusion} Software interface makes it easy for users to write their own (growth-tree damage-economic) models without advanced programming skills. The possibility to run management scenarii with/without considering the impact of natural risk may contribute improving silviculture guidelines and adapting them to climate change. We propose future lines of research and improvement.

%The strong points of the software are its modularity and the  capability for the user to easily write his/her own models without advanced programming skills. The software gives the possibility to run management scenarii with or without considering the impact of natural risk. In the present version, two variables can be taken into account: valuated timber  or sequestered carbon. The optimized criteria used to evaluate the various scenarii are Faustmann and Averaged yield value.  The models, criteria and software structure are discussed and illustrated.
  
  \end{itemize}
\end{abstract}


%%%%%%%%%%%%%%%%%%%%%%%%%%%%%%%%%%%%%%%%%%%%%%%%%%%%%%%%%%%%%%%%%%%%%%%%
\section{Introduction \label{sec:intro}}
 Forest managers have always been interested in the evaluation of silviculture practices. Nowadays, they must consider criteria that cover the  increase in natural risks \citep{Hanewinkel2012} due to climate change. Furthermore,  in the face of growing societal demand,
forest managers must also consider biodiversity, carbon sequestration and more generally ecosystemic
services. For all these criteria, risk must be taken into account and silviculture should evolve accordingly. 

Focusing on forest management at the stand level, a lot of growth models are now available, which 
 were historically dedicated to one or two given
species \citep{Monserud1996, Coates2003, Lacerte2006,    Pretzsch2006}.   Most of them are implemented in software: we do not give details here, but an extensive review can be found in \citet{Capsis}.  Recent efforts have been made to integrate growth models in software platforms in order to fulfill the growing user demand. However existing software suffers from two main limitations : i) it requires a lot of efforts for users to customize the models  used for the simulations ii) it does not integrate natural risks (except ForestGALES for wind risk  \citep{Gardiner2000}, but ForestGALES is very specific and must be coupled with forest growth software). 
 %Based on these growth models and on economic
%criteria, multi-stand planning has been the object of a lot of
%studies.  In order to evaluate multi-stand forest management, it is necessary to
%first have a precise modelling at stand level, before extending it to the multi-level case.

The motivation behind the present work is to answer the limitations of forest simulation software cited above. The challenge is to easily evaluate scenarii while eventually taking  the natural risks into account. We propose a software, called EvaSylv which constitutes a significant advance regarding both these limitations. This software will facilitate the emerging of new silviculture guidelines. %Multi-stand planning is out of the scope of the present work, as it is necessary to first have a precise modelling at stand level, before extending it to the multi-level case.

EvaSylv consists of a chain of interconnected models, including a tree
growth model, a risk model, a tree damage model following risk, and
several economic criteria. The chain of models works at stand level,
and allows to compute analytical expressions for criteria (Faustmann
Value, Averaged Yield Value) linked to a given technical
itinerary. This analytical approach has the advantage to yield precise
results in a short computing time, compared to a Markov-based approach
for computing expected values, which requires a great number of
time-consuming simulations. %{\bf But, by using a procedure based on a
% recurrence equation, EvaSylv offers the possibility to access to the
% distribution of the Faustmann Value, hence the Conditional Value at
% Risk \cite{Rocka} for the Faustmann Value}.

{In EvaSylv, users can easily write their own models through a
  user-friendly interface. The functions and models are considered as
  a supplementary kind of data, and can be modified via the
  interface. Contrary to other forest simulation software, users do
  not have to get into the complex software code. They are written in
  basic Python and do not require advanced programming skills.}

The paper is organized as follows: the framework is introduced in Section \ref{sec:imp}, including a recall of the mathematical models used for simulation and optimization. Software design is presented in Section \ref{design}. Case studies relative to simulation and optimization of beech and pine stands are given in Section \ref{sec:resu}. Section \ref{sec:discu} gives some conclusions and perspectives.

\section{Framework}\label{sec:imp}
The objective of the present work is
to provide an operational easy-to-use simulation and optimization
framework for testing forest management scenarii at stand level.

{EvaSylv has been designed to allow} users to easily write their own models through a user-friendly interface: growth functions, damage models (linked to natural risk), economic valorizations. The functions and models are considered as a supplementary kind of data, and can be modified via the interface, without having to get into the complex software code. They are written in  basic Python and do not require advanced programming skills.

%The software allows to take into account some natural risks during the simulation and optimization processes. 
The evaluation requires the choice of a technical itinerary describing
the authorized actions. For each itinerary, the Faustmann approach is
available to evaluate future benefits, economic or not. It
necessitates the calculation of their expected values, for which a
dedicated mathematical development was written
\citep{Loisel2011,Loisel2014}.

\begin{figure}[hbtp]
\hspace{-2cm}
\includegraphics[width=1.4\textwidth]{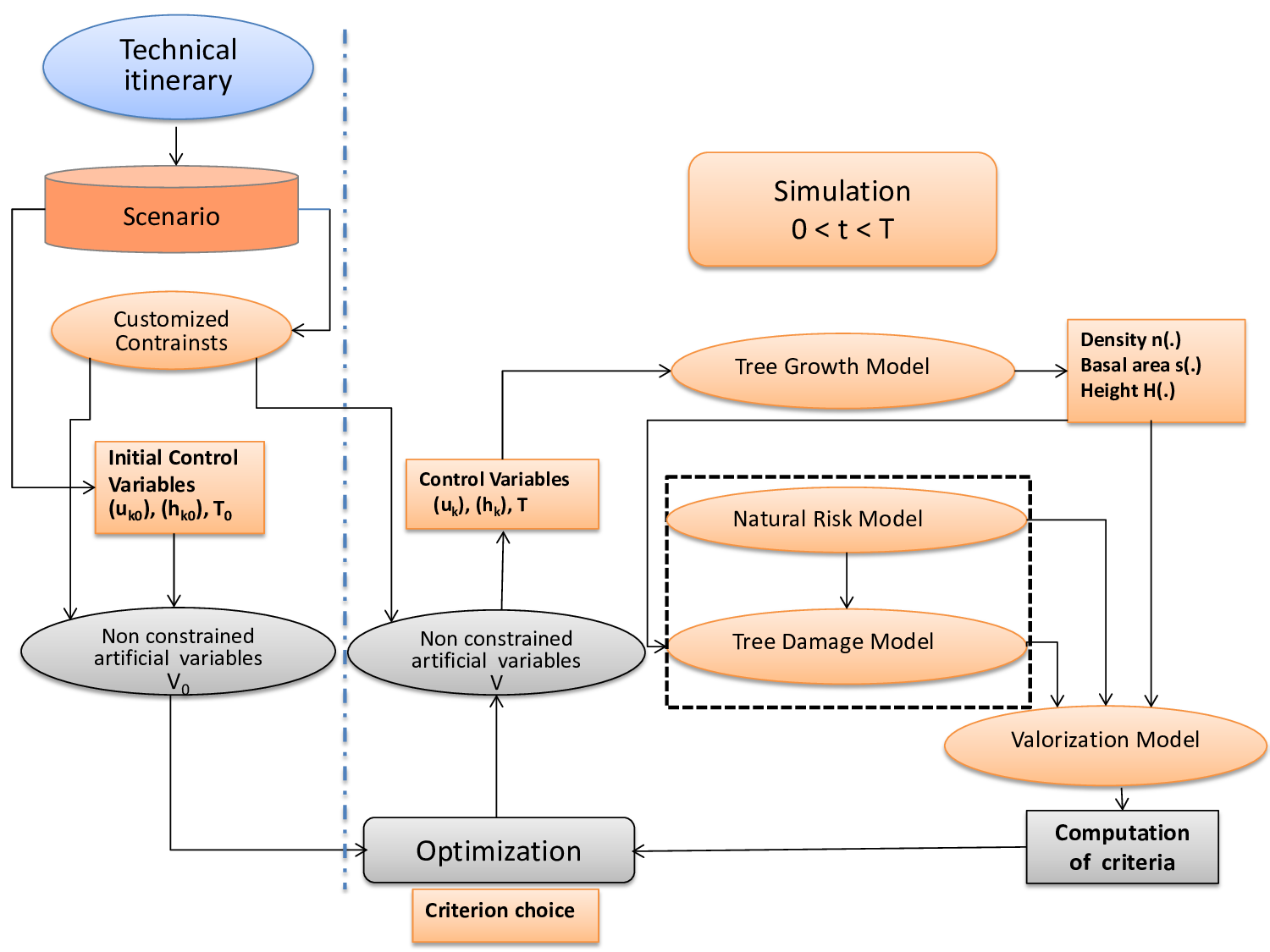}
\caption{Workflow of the proposed modeling approach.}
\label{fig:Gdforet}

\end{figure}
Fig.\ref{fig:Gdforet} represents the workflow of the simulation and
optimization processes within the software.
% Some elements are shaded in gray, while others are within
% orange-colored containers.
{The elements shaded in gray correspond to the in-depth program
  structure and the elements within orange-colored containers
  correspond to the interface}. Particular attention has been paid to
the choice of the latter in order to free the user from writing
complex mathematical calculations whilst ensuring a broad range of
possibilities for designing the scenarii.

\subsection{Software core}\label{sec:meth}
The software core consists of the definition of technical itineraries (with their corresponding scenarii), the set of mathematical models and evaluated criteria, used in the simulation or optimization module.

Different technical itineraries are available, each of them corresponding to a management strategy, taking or not taking the natural risks  into account. Within the technical itinerary, the scenario parameters provide the settings for the initial forest state and for the thinnings: tree-density, dates and rates, as well as the choice of the criterion to optimize, if optimization is required. 
The software can be used in three main modes:
\begin{itemize}
\item Unconstrained simulation: apply a technical itinerary,
\item Constrained simulation : apply an admissible (which satisfies constraints for optimization) technical itinerary,
\item Optimization : find the optimized settings  according to a given criterion.
\end{itemize}
%The first case allows to apply a technical itinerary and a scenario to the entropized forest.  The forest evolution is simulated accordingly . 

The simulation is run over the rotation time $T$. It is based on four interconnected models: tree growth model, natural (such as storm) risk model, tree damage model and gain/cost model. The tree growth model output feeds both the tree damage model and the valorization one. 
The evaluation criteria, based on the Faustmann approach \citep{Faust}, are calculated differently whether or not natural risk is taken into account in the technical itinerary.  In the no risk
case, the criteria are deterministic \citep{Faust}, in case of risk
\citep{Reed84}, they are based on expected values or CVaR.

\subsection{Technical itineraries}\label{sec:scenario}
The technical itineraries constitute a simple framework, based on
production rules
%.  Various technical itineraries are available, each of them having
which its own consequences in terms of income.

In the present version, two technical itineraries are available, the
first one for the no risk case and the second one for the case of
risk. In the no risk case, stand management follows a basic non
ambiguous itinerary: a clearcutting and a regeneration of the stand
are done at the end of rotation period (or cutting age) $T$.  In
presence of natural risk, different technical itineraries may be used
by forest managers.  The software implements an adaptation to storm
risk \citep{Loisel2014} of the itinerary described in \citet{Reed84},
originally designed for forest fire risk and widely used as a
reference in presence of natural risk. It handles storm risk by
defining a rule base, based on the fact that {only} when trees
reach a certain height, storms cause direct damage but also weaken the
trees, increasing future damage. The rule base is as follows:
\begin{itemize}
\item If a storm occurs before Threshold time (user-defined)$t_L$: {no or neglectable impact on
    trees}, then forestry management is unchanged.
\item If a storm occurs at time $\tau$ between $t_L$ and $T$: the ratio of damaged trees is high, then a clearcutting and a regeneration of the stand are performed at time $\tau$.
\item if no storm occurs before time $T$:  a clearcutting and a regeneration of the stand are done at time $T$.
\end{itemize}

In the work presented in this paper, the considered natural risks are
storms, though our approach remains valid for other natural risks,
provided that suited itineraries are elaborated. To each technical
itinerary corresponds associated criteria. {The criteria are of
  different kinds, deterministic or expected actualized value of future
  incomes (Faustmann Value) or expected income (Yield Value),
% valeur aleatoire ou expectation \ref{sec:criteria}, 
  For the expected cases, the expressions of the expectations are
  hard-coded in the software. In order to evaluate the different
  criteria, it is necessary to evaluate incomes and costs. To this
  issue, various models must be given}.

\subsection{A sequence of models}\label{sec:model}

The various model parameters and functions (see Table \ref{tab:not}) are of four different kinds, detailed below:

%L'utilisateur doit fournir les éléments constituant chacun des différents modèles décrivant la croissance, le risque de tempête, les dommages occasionnés par la tempête, les volumes de bois, les gains et coûts. Suivant seuls des paramètres sont demandés ou des fonctions écrites en python. 
%idee : beaucoup de choses à rentrer montre la complexite du pb, pas limite a modele de croissance ``habille''.
\begin{itemize}
\item tree growth model (at stand level). It describes
  the dynamics of the tree basal area $s$: increment growth rate and
  the function describing the evolution of the dominant $H$ and average  $Hmoy$
  tree heights. It is a continuous model which is then discretized.
\item natural risk  model. The natural events occur
  independently of one another, and randomly in time.
% hence the time lapse between two events follows an exponential distribution, of parameter denoted $\lambda$.  $1/\lambda$ represents the expected time lapse between two events.
\item tree damage model. The proportion of damaged trees depends on
  the stand state when the event occurs (in case of storm, see
  \citet{Schmidt2012}). The expected proportion of damaged trees may
  depend on the time, the tree basal area and the tree height.
\item stand valorization model. Two submodels may be considered:
\begin{itemize}
  \item economic model.
%  an economic and an ecosystem service (carbon) model. 
  The economic incomes depend on the timber price functions for
  thinnings and on the final cutting age. 
  \item sequestred carbon model. The sequestred carbon model is related
  to the instantaneous income {and the decreasing following
    thinning or final cut}.
\end{itemize}
\end{itemize}
\begin{center}
\begin{table}[htbp]
\begin{tabular}{|l|l|l|}
\hline
Scope&Variable$/$Parameter&Description\\
\hline\hline
\multirow{3}{22mm}{Growth} & $n$ & tree density (\textit{stems/ha}) \cr
&$s$ & average tree basal area ($m^2$) \cr
&$H_0$ & potential height (linked to fertility) ($m$)\cr 
\hline
{Risk} &{$\lambda$} & risk rate (\textit{year}$^{-1}$) \cr 
\hline
\multirow{2}{22mm}{Tree damage}  &$t_L$ & limit time for tree damage (\textit{year}) \cr
&$H_L$ & limit height for tree damage ($m$) \cr
& $L_s$ & damage rate ($year^{-1}$ \cr
\hline
\multirow{4}{22mm}{Economic parameters}&$\delta$ & discount rate (\textit{year}$^{-1}$) \cr
& $C_n$ & clearing costs after a destructive event (\euro)\cr
&$C_1$ & regeneration cost (\euro)\cr
&$c_w$ & silvicultural cost (\euro)\cr 
&$c_a$ & annual cost (\euro)\cr 
\hline
\multirow{2}{22mm}{Carbon} &$\gamma$ & carbon discount rate (\textit{year}$^{-1}$)\cr
& $Cb$ & sequestred carbon (\textit{ton})  \cr
\hline
\end{tabular}
\caption{Variables and parameters used for simulation.}
\label{tab:not}  
\end{table}
\end{center}
Each model is known with a different accuracy. Growth models are accurate and usually well validated by forestry researchers, while the risk and tree damage models are much more difficult to assess and generally have a low degree of fiability. The economic criteria are usually easy to compute, contrary to the sequestred carbon criteria. It is important to underline that, in a model sequence, the overall accuracy is governed by the less precise model. This justifies the choice made in this first version of EvaSylv to consider tree average-based growth models, instead of modeling basal area structured tree growth {which describes more precisely the tree growth}.

\subsection{Silviculture}
Two phases are to be considered in silviculture. The first phase consists in respacing. It is user defined by a set of two dates and the corresponding respacing rates (Table \ref{tab:silviopt}). The second phase is relative to thinnings. It is also user defined first by the maximum number of thinning dates $N$, then by the thinning dates $(u_k)_{k=1..N}$ and the  corresponding rates of thinnings $(h_k)_{k=1..N}$.  Users must also specify the cutting age $T$.

\begin{table}[htbp]
  \begin{tabular}{|l|l|l|}
    \hline
Scope&Variable$/$Parameter&Description\\
\hline\hline
Respacing &{$h_0, h_1$} & {fixed respacing rates (\textit{year}$^{-1}$)}\cr
&{$td_0, td_1$} & fixed respacing dates (\textit{year})\cr
\hline
\multirow{4}{22mm}{Control variables} &$N$& maximum number of thinning dates \cr
&$h_k$ & $k$th thinning rate (\textit{year}$^{-1}$) \cr
&$u_k$ & $k$th thinning date (\textit{year})\cr
&$T$ & cutting age (free or imposed) (\textit{year})\cr
&$n_f$ & final tree density (constrained or imposed) (\textit{stem/ha})\cr
\hline
\multirow{2}{22mm}{Constraints} &$\Delta$ & minimum elapsed time between thinnings (\textit{year})\cr
&$t_0$ & {reference time for} thinning date (\textit{year}$$)\cr
\hline
\end{tabular}
\caption{Variables and parameters used in silviculture and optimization.}
\label{tab:silviopt}  
\end{table}

\subsection{The criteria used for evaluation}\label{sec:criteria}

{We distinguish the no risk case and the storm risk case.}

{\it (i) In absence of storm risk}:
For a cutting age $T$, a discount rate $\delta$, the Faustmann Value
$J$ taking into account thinning incomes of a stand is the
discounted value of cutting incomes minus cost of regeneration $C_1(n_0)$
and actualized annual costs $c_a$:
\begin{align}
\label{J0}
 J = -{c_a \over \delta} - C_1(n_0)+{W_e(T)-C_1(n_0)\over e^{\delta T} -1}
\end{align}

where $W_e(T)$ is the total economic income on $[0,T]$ composed of the sum of
thinning incomes  $R_k$ at time $u_k$ actualized at time $T$ and the
final income $V(T)$:
\begin{align}
\label{WT}
W_e(T)= \sum_{k=1}^N {R_k.h_k} \ e^{\delta (T-u_k)} + V(T)
\end{align}
For a simplified carbon sequestration model, $J_{C} = {W_c(T) \over e^{\delta T} -1}$ where:
\begin{align}
\label{CT}
W_c(T)= {\delta \over \gamma+\delta} \Big[\sum_{k=1}^N {\ Cb(u_k).h_k} \
e^{\delta (T-u_k)} + Cb(T) \Big] + \delta \int_0^T Cb(u) e^{\delta(T-u)} du
\end{align}

{\it (ii) In presence of storm risk}:

%{\bf La valeur de Faustmann est une variable aleatoire qui peut etre
%  obtenue de facon recurrente}:

%\begin{equation}
%\label{vaF}
%  \mathcal F_i =
%\begin{cases}
%  \displaystyle  e^{-\delta \tau_i} (\sum_{u_k < \tau_i} h_k R_k e^{\delta(\tau_i-u_k)} + (1-L_s(\tau_i)) V(\tau_i)-C_n(\tau_i)-C_1(n_0)+\mathcal F_{i-1})  & \mbox{ if }  t_L < \tau_i < T \\
%  \displaystyle e^{-\delta T} (\sum_{u_k} h_k R_k
%  e^{\delta(\tau_i-u_k)} + V(T)-C_1(n_0)+\mathcal F_{i-1}) & \mbox{ if
%  } \tau_i = T
%\end{cases}
%\end{equation}
%where $\tau_i$ is the spending time between the beginning of the stand
%and the first event of the stand after $t_L$, either by storm or by
%logging at time $T$ for the $i$th rotation. 

%{\bf En considerant l'expectation de l'expression (\ref{vaF}) on
%  deduit classiquement (comme le fait \cite{Reed84} dans le cas sans
%  thinning, \cite{Loisel2014} dans le cas avec thinning)
%  analytiquement l'expression de l'expected Faustmann Value}. 
Let $E_{\mathcal V}(\tau)$ the expected final income and
$E_{C_n}(\tau)$ the expected clearing costs in case of a storm at time
$\tau$. $E_{R_k}$ is the expected potential thinning income at time
$u_k$. The expected Faustmann Value with storm risk becomes:
\begin{align}
\label{Jlambda}
 J_R= E[\mathcal F]= -{c_a \over \delta} - C_1(n_0)+{E_W(T)- C_1(n_0)-C_1(n_0) a(T) \over b(T)} 
\end{align}
where $\displaystyle (\delta+\lambda) a(T) = \lambda
(e^{(\delta+\lambda)(T-t_L)} -1)$, $\displaystyle \ b(T) =
e^{(\delta+\lambda)T-\lambda t_L} - a(T)-1$ and $E_{We}(T)$ is an
expected income, with
\begin{align}
\label{V}
  E_{We}(T) = & \sum_{k=1}^N \beta_{\delta,\lambda}^k {E_{R_k} .h_k} 
%\  e^{(\delta+\lambda)(T-u_k)+\lambda(u_k-t_L)_-}
  +\lambda \int_{t_L}^T [E_{\mathcal V}(\tau)-E_{C_n}(\tau)]
  e^{(\delta+\lambda)(T-\tau)}d \tau + V(T) %\notag
\end{align}
where $\beta^k_{\delta,\lambda} = \beta_{\delta,\lambda}(u_k)$ and 
$ \beta_{\delta,\lambda}(u) = e^{(\delta+\lambda)(T-u)+\lambda(u-t_L)_-}$.

The expressions of the expected values of criteria are hard-coded in
the software. So, the optimization procedure for expected values does
not require simulation of the technical itineraries.

% {\bf Si classiquement dans la litterature citee ci-dessus, on
% s'interesse a l'expectation de la valeur de Faustmann, il peut etre
% pertinent de s'interesser a la distribution
% $\mathcal L_{\mathcal F}$ de la valeur de Faustmann. Pour cela, des
% valeurs de la valeur de Faustmann sont obtenues par simulation via
% la formule de recurrence (\ref{vaF}). De la distribution
% $\mathcal L_{\mathcal F}$, on peut en deduire le criterion
% Conditional Value at Risk ($CVaR(p)$) associated to percentile $p$
% definie comme etant l'expectation de la valeur de Faustmann en
% dessous du quantile associated to percentile $p$}:
%\begin{align}
%  CVAR(p)= & {1 \over p} \int_{\mathcal L_{\mathcal F}(f) < p} f d
%  \mathcal L_{\mathcal F}(f) \notag
%\end{align} {\bf Using the simulated Faustmann values $\mathcal F_l, l
%  = 1..Nsim$, $CVaR(p)$ which can be estimated with simulated
%  Faustmann values: $\displaystyle \widehat{CVaR}(p) =
%  {\sum_{l=1}^{Nsim} \mathcal F_l I_{\mathcal F_l < \widehat{\mathcal
%        F}_p} \over \sum_{l=1}^{Nsim} I_{\mathcal F_l <
%      \widehat{\mathcal F}_p}} $ where empirical quantile
%  $\widehat{\mathcal F}_p$ associated to percentile $p$. Hence:
%  $\displaystyle \widehat{CVaR}(p) = {1 \over N_p} \sum_{l=1}^{N_p}
%  \mathcal F_l$ with $N_p= card\{l|\mathcal F_l < \widehat{\mathcal
%    F}_p \}$.}

For a simplified Carbon sequestration model, we can express a
Faustmann Value $\mathcal F_C$, the corresponding expected Faustmann
value with storm risk $J_{RC}=E[\mathcal F_C]= {E_{Wc}(T)
  \over b(T)}$ where:
\begin{align}
\label{ECT}
E_{Wc}(T) =& {\delta \over \gamma+\delta} \Big[\sum_{k=1}^N
\beta_{\delta,\lambda}^k {Cb(u_k).h_k} +\lambda \int_{t_L}^T
E_{Cb}(\tau) e^{(\delta+\lambda)(T-\tau)}d \tau+
Cb(T)\Big]%/(\gamma+\delta)
\notag\\
&+\delta \int_0^T \beta_{\delta,\lambda}(u) Cb(u) du
\end{align}

In the case of an economic evaluation, the software gives both the
Faustmann criteria and the Long Run Average Yield, with or without
risk. With risk, $CVaR(p)$ is also available.

\subsection{Optimization}
Forest management depends on several control variables (Table
\ref{tab:silviopt}), some of them being scalars, the others being
vectors. The scalar ones include the cutting age $T$ and the final
$n_f$ tree density at time $T$.  The user has the possibility to
optimize the different criteria with the cutting age $T$ imposed or
not, the final tree density $n(T)$ set to $n_f$ or only constrained:
$n(T) \geq n_f$. The respacing dates and rates are control
variables. The vector of control variables is organized in two
sets. The first set is composed of the thinning dates $(u_k)_{k=1..N}$
such that $0 < u_1 < u_2 < .. < u_N < T$ and the second set is
composed of the thinning rates $(h_k)_{k=1..N}$. A minimum time lag
$\Delta$ between two thinnings (starting no sooner than $t_0$) is imposed
in the constrained and optimization modes (see Section
\ref{sec:meth}).

The results are obtained using algorithms to optimize the different
considered criteria. {User selects the choosen algorithm. The
  provided algorithms do not manage constraints on control
  variables}. In order to satisfy constraints, some artificial
variables are introduced (see Appendix A). These artificial variables
are required in both the constrained simulation and optimization
modes, but are not used in the unconstrained simulation mode.

\section{Software design}\label{design}
EvaSylv is designed to let the user set up the runtime behaviour by modifying scalar values and defining custom functions to be used by the mathematical models. Parameters are organized in a hierarchical tree with nodes. A key point is that custom functions are treated exactly as parameters, giving a great flexibility to the models.

\begin{figure}[htbp]
\hspace{-3cm}
\includegraphics[width=1.4\textwidth]{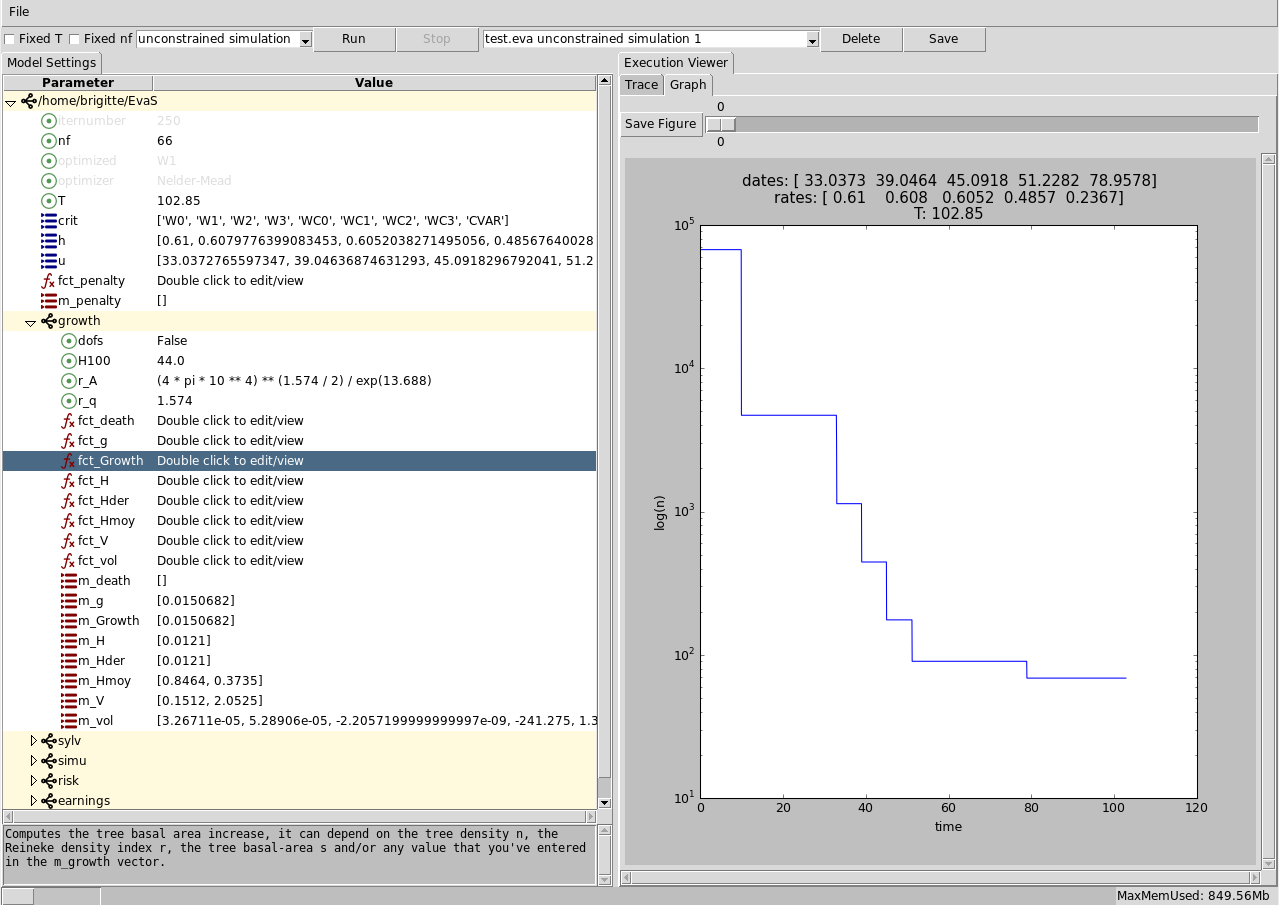}
    \caption{EvaSylv graphical user interface.}
    \label{main}

\end{figure}

In its present version, the program is a stand alone program, distributed in open source license. The software  architecture is schematically composed of three layers %, each of them  appearing in Fig.\ref{scheme}. 
At the deepest level, the \textit{core layer} contains all components of the simulation: model structures and communication between them, technical itineraries and  analytical expressions for the criteria. At the most external level, the \textit{interface layer} handles the  (graphical or command line) interface  that allows to enter a model and its parameters and to build a scenario. In between, the \textit{Build layer} fills in the model structures and scenario parameters for transmission to the \textit{core layer}.
%\begin{center}\begin{figure}[hp]
%\includegraphics[width=\textwidth]{layersEvasylv-red}
%\caption{Software architecture.}
%\label{scheme}
%\end{figure}\end{center}
%\subsection{Programming language}
The software is entirely written in Python. %This language has been chosen for the following reasons :
%\begin{enumerate}
%\item Python is available and easy to install on many operating systems,
%\item It is simple and easy to learn, which reduces the development cost, it has object-oriented programming features and it provides tools to quickly design %a user interface,
%\item It is widely used in the scientific community, so there are very good well tested Python libraries for scientific computation and optimization,
%\item Its syntax is close enough to a mathematical formulation so that it takes little effort for a scientific user, familiar with high level languages such as R or Matlab, to understand how a function is written and to edit it. 
%\end{enumerate}
%\subsection{Practical implementation}
Python version 3.0 has been used, with the main following Python libraries:
\begin{itemize}
    \item \verb+numpy+{(\it www.numpy.org)} provides tools and data structures for high precision
        scientific computation,
    \item \verb+scipy+{\it (www.scipy.org)}, based on \verb+numpy+ provides, among other scientific programming tools, a toolbox for
        optimization,
    \item \verb+sympy+{\it (www.sympy.org)} provides symbolic computation tools.
\item \verb+matplotlib+{\it (www.matplotlib.org)} is used for plotting functionalities.
\end{itemize}
The software can be run in script mode with a configuration file or using a graphical user interface (see screenshot in Fig.\ref{main}). Parameterizing a simulation/optimization is done by double-clicking an item belonging to this tree and editing it. Customizable functions are compiled before the execution of the simulation/optimization functions.

%\subsection{User-defined models}
Users write their own functions using the software interface, which checks syntax validity. These user data are automatically interpreted by the software to build the full model.  Default configurations are provided for beech and pine forests. Users can adapt them to describe other species. 
The most important functions are summarized in Table \ref{tab:model}. The function arguments are limited to the ones given in Table \ref{tab:model}, with an extra parameter vector, as illustrated in \ref{gr} (named $m$ or $p$ in respective functions). 

\begin{table}[h]
\begin{tabular}{|l|l|l|}
\hline
  Models & Purpose& Function\cr
  \hline\hline
\multirow{4}{25mm}{Tree growth} & stand basal area (increase) & $Growth(t,n,s,H_0)$\cr
              & tree height     & $H(t,H_0), Hder(t,H_0),\ Hmoy(t,H_0)$ \cr
              & tree volume     & $vol(t,s,H)$  \cr
              & tree mortality rate & $death(t,n,s)$\cr
\hline
    Tree damage& tree damage rate& $Loss(t,H_L)$ \cr
\hline
  \multirow{2}{25mm}{Economic} & thinnings, final timber price & $Pse(t,s,H),\ Ps(t,s,H)$ \cr
                   & fertilization, clearing costs&$C_f(t),\ C_n(n,Loss,Vol)$         \cr
                   & regeneration, silvicultural costs    & $C_1(n), wc(t,k,n,s,H,Vol)$ \cr
\hline
          Carbon     & {sequestred carbon}    &          $Cb(t,s,H)$ \cr
\hline
\end{tabular}   
\caption{Functions used in models.}
\label{tab:model}         
\end{table}

\section{Results and discussion} \label{sec:resu}
EvaSylv offers the possibility to implement many models, depending on the species and on the taking into account of storm risk, or not. The software functionalities are illustrated with two examples. The first example aims to optimize the economic valorization of a beech stand. The second one is related to the carbon evaluation of a pine stand. In the first case,  silviculture is evaluated for the four types of criteria: Faustmann and Average value with risk and without risk. In the second one, the Faustmann criteria are given.

To enter the sub-models, the user can either use the interface or a configuration file, starting from a default file.

\subsection{Economic evaluation of silviculture for a beech stand}

We consider a beech stand. We are interested in the financial incomes
obtained from the {timber sales}. 

\subsubsection{Growth model}\label{gr}
This sub-model is the one that requires the greatest number of functions. The model core is related to the tree basal area. The tree basal area is governed by a dynamic model, following a differential equation. The user must provide the second member of that equation.
Let us take as example a growth function \citep{Moguedec2012} with one parameter $m_1$ and the following mathematical expression:

\begin{equation*}
(1-e^{-m_1 n(t) \sqrt{4 \pi s(t)}}) V(t,H_0)/n(t)
\end{equation*}
where $V(t,H_0)$ is the potential total basal area increase. \\

The software interface (see Fig. \ref{main}) displays a help text with the growth function purpose and its possible dependencies with other model components, in that case the tree density $n$, the basal area $s$, the potential height $H_0$.  The expression of the growth function is  given below, written in Python code, very close to the mathematical formulation. 

\begin{verbatim}
def fct_Growth(self, t, n, s, m, H0):
    g=((1-exp(-m[1]*n*sqrt(4*pi*s)))*self.fct_V(t,H0))/n
    return g
\end{verbatim}
\color{black}

The growth function definition is prefilled with a default expression. The first line must not be edited. The function takes as argument the $m$ parameter, which is a vector of any length, whose elements will be accessed by using $m[1], m[2]\ldots$ in the function body, which can be edited to fulfill the user needs. {The components of the $m$ vector are specified by the user in a predefined box located above the corresponding function}. The $m$ vector being of variable length, there is a lot of flexibility, as the user can add any parameter of interest. The reference to a pre-defined function must be preceeded by the $self.$ prefix.

The potential total basal area increase is for instance equal to:\\$V(t,H_0)=p_1+p_2 Hder(t,H_0)$, where $\displaystyle
Hder(t,H_0)$ is the derivative of $H(t,H_0)$, the dominant height with
respect to time $t$. 
\begin{verbatim}
def fct_V(self, t, p, H0):
    return p[1]+p[2]*self.fct_Hder(t,H0)
\end{verbatim}
\color{black}
Here is an example of a $H$ function.
\begin{verbatim}
def fct_H(self, t, p, H0):
    return H0 * ( 1 - exp(-p[1] * t))
\end{verbatim}
\color{black}
The derivative is  automatically generated (using  the symbolic computation tools of {sympy}).  $Hder$ is given by:
\begin{verbatim}
def fct_Hder(self, t, p, H0):
    return p[1] * H0 * exp(-p[1] * t)
\end{verbatim}
\color{black}
The user may check it and modify it, if necessary. 

\subsubsection{Optimization}
{Results given in Table \ref{tab:beech} are obtained with the classical Nelder Mead algorithm}. The Faustmann value is calculated with a discount rate $\delta=0.025$ per year. Whatever the chosen criterion, Faustmann value with actualization or Average value, the presence of storm risk reduces the optimal cutting age and therefore the optimal thinning dates occur earlier. This is in agreement with the commonly admitted management strategy. By changing the timber price function $P_s$ (see Table \ref{tab:model}), the user can evaluate the criterion behaviour with respect to the market price.

The introduction of risk in the scenarii has a much greater impact onto the Average value.
This is due to the fact that the loss is increasing with respect to stand age. 

\color{black}
\begin{table}[htbp]
\noindent \begin{tabular}{|ll|cccc|c|c|} \hline
  Criterion & Itinerary &&\multicolumn{3}{c}{Optimal Thinnings} & Cutting age  & Value   \\
% & & \multicolumn{3}{c}{ (year)}& (year)& (Euro/ha)    \\
  \hline\hline
  \multirow{4}{*}{\begin{tabular}{l}Faustmann\\Value\end{tabular}} & \multirow{2}{*}{without risk}   &$u_k$&$47$ & $55$ &$63$&$77.5$ &$2808$ \\
  &&$h_k$& $.249$&$.249$ &$.167$          && \\
  \cline{2-8}
    & \multirow{2}{*}{with risk} &$u_k$& $44$ &$52$&$60$&$68$&$2591$ \\
  & &$h_k$&  $.240$   &$.199$&$.219$& & \\
  \hline
     \multirow{4}{*}{\begin{tabular}{l}Averaged \\Yield Value \end{tabular}} & \multirow{2}{*}{without risk}   &$u_k$& $138$ &$146$ &$154$  &$163$ &$ 313$ \\
     & &$h_k$& $.167$ &$.250$ & $.250$               && \\
     \cline{2-8} 
          & \multirow{2}{*}{with risk}  & $u_k$&$72$ &$80$&$88$ &$96$ & $220$ \\
  &&$h_k$&$.167$&$.250$&$.250$& &\\
  \hline
\end{tabular}
\caption{Optimal Value (\euro/ha) with respect to thinning rates $h_k$, dates $u_k $ and cutting age $T$ (years).}
\label{tab:beech}
\end{table}

The evolution of $n$, the tree density, is plotted in logarithmic
scale {at the right side} in Fig.\ref{main}.
%\begin{figure}[htbp]
%\begin{center}
%\includegraphics[width=0.99\textwidth]{evadensi2}
%    \caption{The tree density variation over time for a beech stand and the Faustmann without risk criterion. Abscissae are in years and ordinates in stems/ha.}
%    \label{fig:n}
%\end{center}
%\end{figure}

\subsection{Evaluation of silviculture for Carbon sequestration of a pine stand} 
We now consider a pine stand and the impact of silviculture on aerial carbon sequestration. The carbon model is currently an open research topic in itself. Many models exist in the literature \citep{Price2011}; \citep{Susaeta201447} with various complexity levels. We chose a simple model, where the carbon function is assumed to be proportional to the tree volume, in order to illustrate the interest of the simulation. 
With or without risk, the criterion to optimize has an integral form (see Eq.\ref{CT} and Eq.\ref{ECT}), so the optimization must be performed with a minimum tree density target.

The software is run with a constraint on the final tree density $n(T)$ , that must be $\geq n_f=200$ stems/ha. Due to the fact that the chosen model for carbon sequestration is an increasing function of time, the optimization will always try to reach the minimum tree density value. The optimal cutting age is found to be $116.4$ years, in both cases, with or without risk. 
The Faustmann criterion value, calculated with a discount rate of $0.03$ per year, is found to be equal to $293$ tons of $CO_2$ for the unrisky case, and $273$ tons of $CO_2$ for the risky case. We arbitrarily assumed that a cubic meter of wood stores a ton of $CO_2$, according to generally admitted values, that range between half a ton and a ton.

%The results shown in Table \ref{tab:pine} confirm this behaviour.

%
%\begin{table}[hbtp]
%\noindent \begin{tabular}{|l|ccccccc|c|} \hline
%  Itinerary &\multicolumn{6}{c}{Optimal Thinnings} & Cutting age  & Value   \\
% & \multicolumn{6}{c}{ (year)}& (year)& (Euro/ha)    \\
%  \hline\hline
%  Faustmann without risk&&&& &&& \\
%  \hspace{8mm}    &&&$47$ & $55$ &$63$&&$77.5$ &$2808$ \\
%  & && $.249$&$.249$ &$.167$          &&& \\
%  &&&&& &&& \\
%  Faustmann with risk &&&&&&& \\
%  \hspace{8mm} && $44$ &$52$&$60$&&&$68$&$2591$ \\
%  &  & $.240$   &$.199$&$.219$&&& & \\
%  \hline
%   Average without risk&&&&&&&& \\
%  \hspace{8mm}   & &&&$138.3$ &$146.4$ &$154.5$  &$162.8$ &$ 312.8$ \\
%  &  & &&$.167$ &$.250$ & $.250$               && \\
%  \hline
%         Average with risk &&&&&&& \\
%  \hspace{8mm}  & $71.5$ &$79.5$&$87.6$ &&&&$95.7$ & $219.8$ \\
%  &$.167$&$.250$&$.250$& &&&&\\
%  \hline
%\end{tabular}
%\caption{Carbon Value with respect to thinning rates $h_k$, dates $u_k $ and cutting age $T$.}
%\label{tab:pine}
%\end{table}

%\input tab.tex

\section{Conclusion}\label{sec:discu}
The EvaSylv software presented in this paper hopes to fill a gap in the available softwares for forest management. 
Even though a lot of forest models are available on tree growth at the stand level, and are included in modeling platforms, they are hard-coded and not user-editable.  EvaSylv gives the users an easy way to simulate their own models at stand level, without needing to have a deep knowledge of computer science languages and of the program structure. This is  the  main strength of the software. EvaSylv functionalities include simulation and optimization tools. Moreover, it allows to take into account natural risks, which is not the case with other forestry simulation platforms. 
 The software relies upon a sequence of integrated models: growth, risk, tree damage and valorization. 

In the present version, limited to storm risk, the software provides two kinds of technical itineraries: a deterministic one (without risk) and a basic one in presence of risk. The modeling in presence of risk requires the calculation of expected values for criteria, which change depending on the technical itinerary. Implementing more complex itineraries requires an interaction with software designers, in order to establish the corresponding evaluation criteria.

The software is likely to participate in the improvement of silviculture guidelines, by taking into account available field data and updated storm risk information.
%Basic functionalities are provided to view and plot the results.

The approach is flexible enough to open several perspectives. First the elaboration of more
complex {technical} itineraries together with forest managers, which implies mathematical developments to obtain the corresponding analytical criteria.
Second the extension to more realistic growth models (structured into classes of basal
area). Third the building of ecosystemic service forest-related models.

\section*{Acknowledgments}
This research received financial support from the Economic, Human and Social Science Researchers Network of Ecofor, (grant number ECOFOR 2012-24). We thank our funders as well as those who provided assistance with writing this paper.
%We would like to thank Hanitra Rakotoarison, from ONF Fontainebleau (France), for having participated in the software testing and validation.
%\section*{Appendix}
\begin{appendix}

\section{Optimization using artificial variables}
The constraints on thinnings dates ($u_k-u_{k-1} \geq \Delta $ years) and, if
necessary, the final tree density constraint ($n(T)=n_f$) are managed
using artificial variables $x_k$ and $y_k$. For exemple, in the case
without mortality:

\begin{align}
  u_k = & u_{k-1} + \Delta + {x_k^2 \over \sum_{j=1}^N x_j^2 +1}
  |x_{N+1}|,&  k = 1..N \notag \\
  T = & t_0+\Delta. N+|x_{N+1}|,&  \notag \\
  h_k = & 1 - ({n_f \over n_0(1-h_0)(1-h_1)})^{y_k^2 \over \sum_{j=1}^N
    y_j^2},& k = 1..N \notag
\end{align}

{Although the Nelder Mead algorithm has a better behavior than the
  steepest descent method, it is likely to get stuck in local
  minima (maxima in our work). The implemented algorithm does not
  ensure to obtain the solution corresponding to the global maximum. To avoid this problem, it
  is recommended to run the optimization with several initial
  conditions.}

The automatic transformation back and forth between artificial and real control variables is implemented in the software workflow, as represented in Fig.\ref{fig:Gdforet}.
\end{appendix}

%\bibliography{}

\begin{thebibliography}{}

\end{thebibliography}


\begin{thebibliography}{}
\bibitem[Coates et~al., 2003]{Coates2003}
Coates, K., Canham, C.~D., Beaudet, M., Sachs, D.~L., and Messier, C. (2003).
\newblock Use of a spatially explicit individual-tree model (sortie/bc) to
  explore the implications of patchiness in structurally complex forests.
\newblock {\em Forest Ecology and Management}, 186(1-3):297 -- 310.

\bibitem[Dufour-Kowalski et~al., 2012]{Capsis}
Dufour-Kowalski, S., Courbaud, B., Dreyfus, P., Meredieu, C., and Coligny, F.
  (2012).
\newblock Capsis: an open software framework and community for forest growth
  modelling.
\newblock {\em Annals of Forest Science}, 69(2):221--233.

\bibitem[Faustmann, 1849]{Faust}
Faustmann, M. (1849).
\newblock Berechnung des wertes welchen waldboden sowie noch nicht haubare
  holzbest{\"a}nde f{\"u}r die waldwirtschaft besitzen.
\newblock {\em Allgemeine Forst-und Jagd-Zeitung}, 15(1849):7--44.

\bibitem[Gardiner and Quine, 2000]{Gardiner2000}
Gardiner, B.~A. and Quine, C.~P. (2000).
\newblock Management of forests to reduce the risk of abiotic damage - a review
  with particular reference to the effects of strong winds.
\newblock {\em Forest Ecology and Management}, 135(1-3):261--277.

\bibitem[Hanewinkel et~al., 2012]{Hanewinkel2012}
Hanewinkel, M., Cullmann, D.~A., Schelhaas, M.-J., Nabuurs, G.-J., and
  Zimmermann, N.~E. (2012).
\newblock Climate change may cause severe loss in the economic value of
  european forest land.

\bibitem[Lacerte et~al., 2006]{Lacerte2006}
Lacerte, V., Larocque, G., Woods, M., Parton, W., and Penner, M. (2006).
\newblock Calibration of the forest vegetation simulator (fvs) model for the
  main forest species of ontario, canada.
\newblock {\em Ecological Modelling}, 199(3):336 -- 349.

\bibitem[Le~Moguedec and Dhote, 2012]{Moguedec2012}
Le~Moguedec, G. and Dhote, J.-F. ({2012}).
\newblock {Fagacees: a tree-centered growth and yield model for sessile oak
  (Quercus petraea L.) and common beech (Fagus sylvatica L.)}.
\newblock {\em {Annals of Forest Science}}, {69}({2}):{257--269}.

\bibitem[Loisel, 2011]{Loisel2011}
Loisel, P. (2011).
\newblock Faustmann rotation and population dynamics in the presence of a risk
  of destructive events.
\newblock {\em Journal of Forest Economics}, 17(3):235 -- 247.

\bibitem[Loisel, 2014]{Loisel2014}
Loisel, P. (2014).
\newblock Impact of storm risk on faustmann rotation.
\newblock {\em Forest Policy and Economics}, 38(0):191--198.

\bibitem[Monserud and Sterba, 1996]{Monserud1996}
Monserud, R.~A. and Sterba, H. (1996).
\newblock A basal area increment model for individual trees growing in even-
  and uneven-aged forest stands in austria.
\newblock {\em Forest Ecology and Management}, 80(1-3):57 -- 80.

\bibitem[Pretzsch et~al., 2006]{Pretzsch2006}
Pretzsch, H., Biber, P., Iursky, J., and Sodtke, R. (2006).
\newblock The individual-tree-based stand simulator silva.
\newblock In Hasenauer, H., editor, {\em Sustainable Forest Management}, pages
  78--84. Springer Berlin Heidelberg.

\bibitem[Price and Willis, 2011]{Price2011}
Price, C. and Willis, R. (2011).
\newblock The multiple effects of carbon values on optimal rotation.
\newblock {\em Journal of Forest Economics}, 17(3):298 -- 306.

\bibitem[Reed, 1984]{Reed84}
Reed, W.~J. (1984).
\newblock The effects of the risk of fire on the optimal rotation of a forest.
\newblock {\em Journal of Environmental Economics and Management}, 11(2):180 --
  190.

%\bibitem[Rockafellar and Uryasev, 2000]{Rocka}
%Rockafellar, R. T. and Uryasev, S. (2000).
%\newblock Optimization of conditional value-at-risk. 
%\newblock {\em Journal of risk}, 2, 21-42.

\bibitem[Schmidt et~al., 2010]{Schmidt2012}
Schmidt, M., Hanewinkel, M., Kandler, G., Kublin, E., and Kohnle, U. (2010).
\newblock An inventory-based approach for modeling single-tree storm damage
  experiences with the winter storm of 1999 in southwestern germany.
\newblock {\em Canadian Journal of Forest Research}, pages 1636--1652.

\bibitem[Susaeta et~al., 2014]{Susaeta201447}
Susaeta, A., Chang, S.~J., Carter, D.~R., and Lal, P. (2014).
\newblock Economics of carbon sequestration under fluctuating economic
  environment, forest management and technological changes: An application to
  forest stands in the southern united states.
\newblock {\em Journal of Forest Economics}, 20(1):47 -- 64.

\end{thebibliography}
%\bibliographystyle{elsarticle-num-names}
%\bibliographystyle{plainnat}
\bibliographystyle{apalike}

\end{spacing}
\end{document}